\documentclass[11pt]{article} 
\usepackage{hyperref} 
\pdfoutput=1

\begin{document}

\title{The structure and dynamics of self-assembling colloidal monolayers in oscillating magnetic fields}

\author{Alison E. Koser, Nathan C. Keim, and Paulo E. Arratia}

\maketitle
\begin{center}
Department of Mechanical Engineering and Applied Mechanics\\
 University of Pennsylvania, Philadelphia, USA\\%

\end{center}

Many fascinating phenomena such as large-scale collective flows, enhanced fluid mixing and pattern formation have been observed in so-called active fluids, which are composed of particles that can absorb energy and dissipate it into the fluid medium. Examples of active fluids include swarming bacteria and early embryonic development. In order to investigate the role of hydrodynamic interactions in the collective behavior of an active fluid, we choose a model system: paramagnetic particles submerged in water and activated by an oscillating magnetic field. The particles are $20 \mu$m in diameter, and the oscillating magnetic field (frequency 0.2 Hz) is generated via four computer-controlled electromagnets. The magnetic field induces magnetic attractions among the paramagnetic particles, activating the particles, and injecting energy into the fluid. Over many cycles, the particles aggregate together and form clusters. In order to form clusters, however, the particles must overcome viscous drag. We investigate the relative roles of viscosity and magnetism. When the role of viscosity is important, the particles cannot form large clusters. But when the role of magnetism is important, the particles rapidly form organized, large clusters. Our results shown in this fluid dynamics video suggest that viscous stresses slow the clustering rate and decrease the size of clusters in a self-assembling colloidal system.





\end{document}